\documentclass[11pt]{article}
\def\s{\sigma} 
\def\o{\omega} 

\def\r{\ref}
\def\p{\partial}
\def\no{\nonumber}
\begin{document}
\title{{\bf{\Large  Normal ordering and non(anti)commutativity 
\\in open super strings
 }}}
\author{
 {\bf {\normalsize Sunandan Gan{g}opadhyay}$^{a,}
$\thanks{sunandan@bose.res.in}},\, 
 {\bf {\normalsize Arindam Ghosh Hazra}$^{a,b,}
$\thanks{arindamg@bose.res.in}}\\
$^{a}$ {\normalsize S.~N.~Bose National Centre for Basic Sciences,}\\{\normalsize JD Block, Sector III, Salt Lake, Kolkata-700098, India}\\[0.3cm]
$^{b}$ {\normalsize Sundarban Mahavidyalaya}\\{\normalsize 
Kakdwip, South 24 Parganas, West-Bengal, India}\\[0.3cm]
}

\maketitle

\begin{abstract}
Nonanticommutativity in an open super string moving in the presence of a
background antisymmetric tensor
field $\mathcal{B}_{\mu \nu}$ is investigated in a conformal 
field theoretic approach, leading
to  nonanticommutative structures. In contrast to several 
discussions, in which boundary conditions are taken as Dirac 
constraints, we first obtain the mode algebra by using the newly
proposed normal ordering, which satisfies both equations of motion and boundary 
conditions. Using these
the anticommutator among the fermionic string coordinates is obtained.
Interestingly, in contrast to the bosonic case,
this new normal ordering  plays an important role in uncovering the underlying
nonanticommutative structure between the fermionic string coordinates.  
We feel that our approach is more transparent than the previous ones and the 
results we obtain match 
with the existing results in the literature.
\\[0.3cm]

\end{abstract}

\section{Introduction}
Recent progress in string theory \cite{maldacena, randall, sundrum} indicates
 scenarios
where our four dimensional space-time with standard model fields 
corresponds to a D3-brane \cite{polchinski} embedded in a larger manifold.
Now, since D-branes correspond, in type II string theories, to the space
where the open string endpoints are attached, our space-time would be affected
by string boundary conditions (BC). One important consequence is the possible
noncommutativity of space-time coordinates at very small length scales
\cite{chu, schomerus, witten} since commuting coordinates are incompatible
 with open string
BC(s) in the presence of antisymmetric tensor backgrounds. This is one of the 
main reasons of increasing interest in several aspects of noncommutative (NC)
quantum field theories \cite{witten, szabo}. Furthermore,
this illustrates the fact
that the string BC(s) may play a crucial role in the
phenomenology of four-dimensional physics.

\noindent Various methods have been applied to obtain this result
\cite{chu, chu1, ardalan, rb, agh, jing, our, sg}.
One of the most conventional methods to derive this noncommutativity
is to use Dirac'procedure \cite{dirac}-treat the mixed BC(s)
as primary constraints \cite{shirzad}. However, the interpretation of the
BC(s) as primary Dirac constraints lead to some ambiguities. One of them is,
in contrast to the standard Dirac's method, one obtains infinite secondary
constraint chains by consistency requirements \cite{shirzad, deghani}.
The other is that
the BC(s) are valid only at the boundaries, the Dirac function is introduced
so as to extend them to the neighborhood of the boundaries. In order to get
a finite result, one should regularise the Dirac function.
However, different regularisation procedures may lead to different
results \cite{loran}; a regularisation free proof is needed.

\noindent In a series of recent papers involving bosonic strings \cite{rb}
and superstrings \cite{agh},
it has also been shown explicitly
that noncommutativity can be obtained by modifying the canonical 
bracket structure, so that it is compatible with the BC(s). 
This is done in spirit to 
the treatment of Hanson et.al \cite{hrt}, where modified Poisson brackets 
(PBs) were obtained for the free Nambu-Goto string.
In \cite{jing}, PBs among the Fourier components were obtained using the
Faddeev-Jackiw symplectic formalism \cite{fj}, so that they are compatible
with these BCs. Using this the PBs
among the open string coordinates were computed revealing the NC structure
in the string end- points. It is important to note that all these analysis
were essentially confined to the classical level.
In a very recent paper \cite{our}, noncommutativity in an open bosonic string
moving in the presence of a
background Neveu-Schwarz two-form
field $B_{\mu \nu}$ is investigated in a conformal 
field theory approach. The mode algebra is first obtained using the newly
proposed normal ordering \cite{godinho}, which satisfies both
equations of motion and BC(s). Using these
the commutator among the string coordinates is obtained. Interestingly, 
this new normal ordering  yields the same algebra 
between the modes as the one satisfying only  the equations of motion. 
In this approach, we find that noncommutativity originates more 
transparently and our results match with the existing results
in the literature.
In this paper, we shall extend the methodology of \cite{our} to
analyse an open super string propagating freely and one moving 
in a constant antisymmetric background field.

\noindent The organisation of this paper is as follows. 
In section 2, we review 
the recent results involving new normal ordered products
(of fermionic operators) in \cite{br}.
In sec 3, we study the symplectic structure of the fermionic sector of 
both free and interacting
super string using the method in \cite{our}
(borrowing relevant results from section 2).
We conclude in section 4. The computational details of some of the
key results in the paper are given in an appendix.


\section{New Normal ordering for fermionic string coordinates}
The action for a super string moving in the presence of 
a constant background antisymmetric tensor field 
${\cal B}_{\mu \nu}$ is given by:
\begin{eqnarray}
S &=&  \frac{- 1}{4 \pi \alpha^\prime} \int_{\Sigma} d\tau
 d{\s}\,\Big[ \, \partial_a X^\mu \partial^a X_\mu 
\,+\, \epsilon^{ab} B_{\mu\nu} \partial_a X^\mu \partial_b X^\nu
\nonumber\\
 & & + i \psi_{\mu (-)} E^{\nu \mu} \partial_{+} \psi_{\nu (-)} +
i \psi_{\mu (+)} E^{\nu \mu} \partial_{-} \psi_{\nu (+)} \,\Big]
\label{1}
\end{eqnarray}
\noindent where, $\partial_{+} =  \partial_{\tau} + \partial_{\s},\; \;
\partial_{-} =  \partial_{\tau} - \partial_{\s} $
and $ E^{\mu\nu} \,=\, \eta^{\mu\nu} \, + {\cal B}^{\mu\nu}\,$.

\noindent Now since the bosonic and fermionic sectors decouple, we 
can consider the fermionic sector seperately\footnote{The bosonic sector 
was already discussed in \cite{our}.}. 
The variation of the fermionic part of the action (\r{1})
gives the classical equations of motion: 
\begin{equation}
\partial_{+} \psi_{\nu (-)} \,=\, 0\quad,\quad
\partial_{-} \psi_{\nu (+)} \,=\, 0
\label{2}
\end{equation}
and a boundary term that yields the following BCs:
\begin{eqnarray}
 E_{\nu\mu}\,  \psi^\nu_{(+)} (0,\tau)\, &=&\,
  E_{\mu\nu} \, \psi^\nu_{(-)} (0,\tau)\, \nonumber\\
\label{4}
E_{\nu\mu}\,  \psi^\nu_{(+)} (\pi,\tau ) \, &=& \,\lambda
E_{\mu\nu} \, \psi^\nu_{(-)} (\pi, \tau )
\label{3}
\end{eqnarray}
\noindent at the endpoints $\s \,=\,0$ and $\s = \pi\,$ of the string,
where $\lambda = \pm 1 \,$ corresponds to Ramond and Neveu-Schwarz BC(s) 
respectively.

It is convenient now to change to complex world-sheet coordinates and 
therefore we first
make a Wick rotation by defining
$\s^2 = i\tau$. Then we introduce the complex world 
sheet coordinates \cite{pol}: 
$ z \,=\,  \s^1 +i \s^2 \, ; \,  \bar{z} \,=\, \s^1 - i \s^2$
and $\p_z = \frac{1}{2}(\p_1 - i\p_2), \, \p_{\bar{z}} = 
\frac{1}{2}(\p_1 + i\p_2)$. In this notation the fermionic part of the 
action (\ref{1}) reads:
\begin{eqnarray}
S_{F} =  \frac{-i}{4 \pi \alpha^\prime} \int_{\Sigma} dz
d \bar{z}[\psi_{\mu (-)} E^{\nu \mu} \partial_{\bar{z}} \psi_{\nu (-)} +
\psi_{\mu (+)} E^{\nu \mu} \partial_{z} \psi_{\nu (+)}]\,
\label{5}
\end{eqnarray}
while the classical equations of motion (\ref{2}) and the BCs (\ref{3})
take the form:
\begin{eqnarray}
\label{6}
\partial_{\bar{z}} \psi_{\nu (-)} \,=\, 0\,\,\,, \,\,
\partial_{z} \psi_{\nu (+)} \,=\, 0 \\
\left( E_{\nu\mu}\,  \psi^\nu_{(+)} (z,\bar{z})\, -\,
  E_{\mu\nu} \, \psi^\nu_{(-)} (z,\bar{z})\right)\vert_{z\,=\,
-\bar{z}\, , \, 2\pi - \bar{z}}\, =\, 0 \, .
\label{7}
\end{eqnarray}
We now study the properties of quantum operators
corresponding to the classical variables by considering the 
expectation values\cite{pol}. Using the fact that the path integral of a 
total functional derivative vanishes and considering the insertion of one 
fermionic operator one finds:
\begin{equation}
\label{8}
\int [ d  \psi ] \left[  \frac{\delta}{\delta \psi^{\mu} _{(a) }(z,\bar{z})} 
[e^{-S_{F}} \psi^{\nu}_{(b)} (z', \bar{z}')]
  \right] \,=\,0
\end{equation}
\noindent where, $a,b \,=\,+,-\,$.
Considering first the case of 
$\psi^{\nu}_{(b)} (z', \bar{z}')\,$ inside the world-sheet 
and not at the boundary,
this equation yields the following expectation values:
\begin{eqnarray}
\label{eqmov}
\langle \partial_{z} \psi^{\mu}_{(+)} (z,\bar{z}) \psi^{\nu}_{(+)}
(z',\bar{z}') \rangle &=& 2\,\pi \,i \, \alpha'\, \langle \eta^{\mu \nu} 
\delta^2(z-z',\bar{z}-\bar{z}') \rangle \nonumber \\
\langle \partial_{\bar{z}} \psi^{\mu}_{(-)} (z,\bar{z})
\psi^{\nu}_{(-)}(z',\bar{z}') \rangle &=& 2\,\pi \,i \, \alpha '\, \langle
 \eta^{\mu\nu}  \delta^2(z-z',\bar{z}-\bar{z}') \rangle \nonumber \\
\langle \partial_{\bar{z}} \psi^{\mu}_{(-)} (z,\bar{z})
\psi^{\nu}_{(+)}(z',\bar{z}') \rangle &=&  \langle
 \partial_{z} \psi^{\mu}_{(+)} (z,\bar{z})
\psi^{\nu}_{(-)}(z',\bar{z}') \rangle =  0\,\,.
\end{eqnarray}
Using these results one finds the appropriate way to 
define normal ordered products
that satisfy the equations of motion
for fermionic operators that are not at the world-sheet 
boundary\cite{pol1, br}:   
\begin{eqnarray}
\label{9}
: \psi^{\mu}_{(+)}(z,\bar{z})\,\,\, \psi^{\nu}_{(+)}(z',\bar{z}')  : &=& 
  \psi^{\mu}_{(+)}(z,\bar{z})\,\, \,\psi^{\nu}_{(+)}(z',\bar{z}') 
\,-\, \frac{i\, \alpha'\,}{{\bar z} - {\bar z}'}\,\eta^{\mu \nu}
\nonumber\\
: \psi^{\mu}_{(-)}(z,\bar{z})\,\,\, \psi^{\nu}_{(-)}(z',\bar{z}')  : &=& 
  \psi^{\mu}_{(-)}(z,\bar{z})\,\,\, \psi^{\nu}_{(-)}(z',\bar{z}') 
\,-\, \frac{i\, \alpha'\,}{ z -  z'}\,\eta^{\mu \nu}
\nonumber\\
: \psi^{\mu}_{(+)}(z,\bar{z})\,\,\, \psi^{\nu}_{(-)}(z',\bar{z}')  : &=& 0
\nonumber\\
: \psi^{\mu}_{(-)}(z,\bar{z})\,\,\, \psi^{\nu}_{(+)}(z',\bar{z}')  : &=& 0\,.
\end{eqnarray}
The above products satisfy the equations of motion (\r{6})
at the quantum level, 
but fails to satisfy the BC(s) (\r{7}).

\noindent At this point it is more convenient to choose world sheet 
coordinates, related to these $z$ coordinates by conformal
transformation, that simplify the representation of the boundary,
\begin{eqnarray}
\o\, = \,\mathrm{exp}\left(-iz\right)\, =\, e^{- i \s^1 + \s^2} \,\,\,;\,\,\,{\bar \o} = e^{i \s^1 + \s^2}. 
\label{10}
\end{eqnarray}
Besides replacing 
$\mathrm{exp}(-iz) \to \o$, we must transform the fields \cite{pol1},
\begin{eqnarray}
\label{13}
\psi^{\mu}_{\o^{\frac{1}{2}}}(\o) \,=\, \left(\p_\o z\right)^{\frac{1}{2}}
\psi^{\mu}_{z^{\frac{1}{2}}}(z)
= \, i^{\frac{1}{2}} \o^{-\frac{1}{2}}\, \psi^{\mu}_{z^{\frac{1}{2}}}(z).
\end{eqnarray}
The subscripts are a reminder that these transform with half the weight
of a vector. 
In this present coordinates the complete boundary corresponds 
just to the region $\o = {\bar \o}$.
Further, the action (\r{5})  along with equations of motion (\r{6})
in terms of $\o, {\bar \o}$ has still the same form, while the form of BC(s)
(\ref{7}) change to the following:
\begin{eqnarray}
\left( E_{\nu\mu}\,  \psi^\nu_{(+)} (\o,\bar{\o})\, + i\,
  E_{\mu\nu} \, \psi^\nu_{(-)} (\o,\bar{\o})\right)\vert_{\o\,
=\,\bar{\o}}\, =\, 0 \, .
\label{28}
\end{eqnarray}

\noindent Let us now consider the case of an insertion of a fermionic 
string coordinate $\psi^{\nu}_{(\pm )} (\o')\,$ 
located at the world-sheet boundary. Note that since $\o' = 
\bar{\o}'$ at the boundary, the fermionic coordinate insertion at
the boundary depends only on $\o'$. 
Working out equation 
(\ref{8}), but now subject to constraint (\ref{28}) (with $\o$ replaced by
$\o'$ in (\ref{28})), we find\footnote{Note that the fields $\psi$'s with unprimed arguements are not 
located at the boundary.} (see appendix for the computational details):
\begin{eqnarray}
\label{eqmov2}
\langle \partial_{\o} \psi^{\mu}_{(+)} (\o,\bar{\o}) \psi^{\nu}_{(+)}
(\o') \rangle &=& \, 2\pi i\alpha^{\prime}\langle \eta^{\mu \nu} 
\delta^2(\o-\o',\bar{\o}-\o') \rangle \nonumber \\
\langle \partial_{\o} \psi^{\mu}_{(-)} (\o,\bar{\o}) \psi^{\nu}_{(-)}
(\o') \rangle &=& \, 2\pi i\alpha^{\prime}\langle \eta^{\mu \nu} 
\delta^2(\o-\o',\bar{\o} - \o') \rangle \nonumber \\
\langle \partial_{\o} \psi^{\mu}_{(+)} (\o,\bar{\o}) \psi^{\nu}_{(-)}
(\o') \rangle &=& \, -2\pi \alpha^{\prime}
\langle\Big[ ( \eta+{\cal B})^{^{-1}}\,
( \eta - {\cal B}) \Big]^{\nu \mu} 
\delta^2(\o-\o',\bar{\o}-\o') \rangle \nonumber \\
\langle \partial_{\o} \psi^{\mu}_{(-)} (\o,\bar{\o}) \psi^{\nu}_{(+)}
(\o') \rangle &=& \, 2\pi \alpha^{\prime}\langle
\Big[ ( \eta+{\cal B})^{^{-1}}\,
( \eta - {\cal B}) \Big]^{\nu \mu} 
\delta^2(\o-\o',\bar{\o}-\o') \rangle \,\,.
\end{eqnarray}
So the appropriate normal ordering for fermionic string 
coordinates at the boundary reads:
\begin{eqnarray}
:\psi^{\mu}_{(+)} (\o,\bar{\o}) \psi^{\nu}_{(+)}
(\o') : &=&  \psi^{\mu}_{(+)} (\o,\bar{\o}) \psi^{\nu}_{(+)}
(\o')  \,-\, \frac{i\alpha^{\prime}}{\left(\bar{\o} - \o'
\right)} \eta^{\mu\nu}
\nonumber\\
: \psi^{\mu}_{(-)} (\o,\bar{\o}) \psi^{\nu}_{(-)}
(\o') : &=& \psi^{\mu}_{(-)} (\o,\bar{\o}) \psi^{\nu}_{(-)}
(\o')  \,-\, \frac{i\alpha^{\prime}}{\left(\o - \o'\right)}
\eta^{\mu\nu}
\nonumber\\
:  \psi^{\mu}_{(+)} (\o,\bar{\o}) \psi^{\nu}_{(-)}
(\o')  : &=&   \psi^{\mu}_{(+)} (\o,\bar{\o}) \psi^{\nu}_{(-)}
(\o') \,+\,\frac{ \alpha^{\prime}  
\, \Big[ ( \eta + {\cal B})^{^{-1}}\,( \eta  - 
{\cal B}) \Big]^{\nu \mu}}{\left(\bar{\o} - \o'\right)} 
\nonumber\\
:  \psi^{\mu}_{(-)} (\o,\bar{\o}) \psi^{\nu}_{(+)}
(\o')  : &=&  \psi^{\mu}_{(-)} (\o,\bar{\o}) \psi^{\nu}_{(+)}
(\o') \,-\,\frac{ \alpha^{\prime}  
\, \Big[ ( \eta + {\cal B})^{^{-1}}\,(\eta  - {\cal B}) \Big]^{\nu 
\mu} }{\left(\o - \o'\right)}
\label{11}
\end{eqnarray}
The above results of normal ordering of fermionic operators are new and
incorporates the effect of BC(s).

\noindent Now for the functional ${\cal{F}}[X]$ (representing
the combinations occuring in the left hand side of the above
equation), the new 
normal ordering (in absence of the $\mathcal{B}$ field)
can be compactly written as:
\begin{eqnarray}
\label{compact}
:{\cal{F}}:= \exp\left(
\frac{i\alpha^{\prime}}{2}
\int d^2 \o^{\prime\prime}
d^2 \o^{\prime\prime\prime}\left[\frac{1}{(\omega^{\prime\prime}-
\omega^{\prime\prime\prime})}\frac{\delta}{\delta \psi^{\mu}_{(-)}
(\omega^{\prime\prime}, \bar{\omega}^{\prime\prime})}\,
\frac{\delta}{\delta\psi_{\mu(-)}
(\omega^{\prime\prime\prime}, \bar{\omega}^{\prime\prime\prime})}+\left(\o\leftrightarrow\bar\o,
-\leftrightarrow +\right)\right]\right)\cal{F}\no\\
\end{eqnarray}
Note that the fields $\psi$'s with double prime and triple
prime arguements in (\ref{compact}) are not located at the boundary.

\noindent
We shall see now that normal ordered products are important to
compute the central charge which gives us the critical dimension.
\noindent The energy-momentum tensor 
(in the absence of the $\mathcal{B}$ field) for the fermionic sector
for points inside the world-sheet (in the $z$-frame) is given by:
\begin{eqnarray}
\label{energy}
T^{zz}=-\frac{1}{2}\psi_{\mu(+)}\partial_{\bar{z}}\psi^{\mu}_{(+)}
\equiv\bar{T}\nonumber\\
T^{\bar{z}\bar{z}}=-\frac{1}{2}\psi_{\mu(-)}\partial_{z}\psi^{\mu}_{(-)}
\equiv T
\end{eqnarray}
while at the boundary, the BC(s) (\ref{7}) (with $\mathcal{B}=0$)
relating $\psi_{\nu(-)}$ to $\psi_{\nu(+)}$ lead to
:
\begin{eqnarray}
\label{energy1}
\bar{T}=-\frac{1}{2}\psi_{\mu(+)}\partial_{\bar{z}}\psi^{\mu}_{(+)}=-T
\end{eqnarray}
where we have used $\partial_{\bar{z}}=-\partial_{z}$ (since 
$dz=-d\bar{z}$ at the boundary).
The central charge can now be computed from the most singular term
in the normal ordered product of energy-momentum tensor. 
This involves two contractions of the fermionic coordinate operator 
products and is proportional to \cite{br}:
\begin{eqnarray}
\label{energy2}
\int dz^{\prime}...dz^{\prime\prime\prime\prime}\frac{1}{2}
\left[\frac{i\alpha^{\prime}}{(z^{\prime}-z^{\prime\prime})}
\frac{\delta^{2}}{\delta\psi_{\mu(-)}(z^{\prime})
\delta\psi^{\mu}_{(-)}(z^{\prime\prime})}\right]
\left[\frac{i\alpha^{\prime}}{(z^{\prime\prime\prime}-
z^{\prime\prime\prime\prime})}
\frac{\delta^{2}}{\delta\psi_{\mu(-)}(z^{\prime\prime\prime})
\delta\psi^{\mu}_{(-)}(z^{\prime\prime\prime\prime})}\right] \no \\
\times \left[T(z_{1})T(z_{2})\right]\no \\
\sim \frac{D\alpha^{\prime 2}}{4(z_{1}- z_{2})^{4}}\quad \quad \quad
\quad \quad \quad \quad \quad \quad \quad \quad \quad \quad \quad \quad \quad \quad \quad \quad \quad \quad \quad \quad \quad \quad \quad 
\quad \quad \quad \quad 
\end{eqnarray}
where $\sim$ mean ``equal up to nonsingular terms"\footnote{The other
less singular terms are not given explicitly.}. 
The above computation gives the well known result
$D/2$ as the central charge where $D$ is the dimension of space-time 
\cite{pol1}, \cite{kaku}.
The results are also in conformity with \cite{br}.

\noindent We shall make use of the results discussed here in the next section
where we study both free and interacting open super strings.
\section{Mode expansions and Non(anti)Commutativity for super strings} 
\subsection{Free open strings} 
In this section, we consider the mode expansions of free 
(${\cal B}_{\mu \nu} = 0$) open super strings.
We first expand $\psi^{\mu}_{(-)}(z)$ and 
$\psi^{\mu}_{(+)}(\bar{z})$ in Fourier modes in ($z, \bar{z}$) 
coordinates\cite{pol1}:
\begin{eqnarray}
\label{12}
\psi^{\mu}_{(-)}(z) = \frac{1}{\sqrt{2\pi}}\sum_{m \in {\bf{Z}}} 
d^{\mu}_{m}\,\mathrm{exp}(imz)\ \ \ \ ;\ \ \ \ 
\psi^{\mu}_{(+)}(\bar{z}) = \frac{1}{\sqrt{2\pi}} \sum_{m \in {\bf{Z}}} 
\tilde{d}^{\mu}_{m}\,\mathrm{exp}(-im\bar{z}).
\end{eqnarray}
Let us also write these as Laurent expansions in ($\o, \bar{\o}$) coordinates:
\begin{eqnarray}
\label{14}
\psi^{\mu}_{(-)}(\o) = \frac{i^{\frac{1}{2}}}{\sqrt{2\pi}}
\,\sum_{m \in {\bf{Z}}}
\frac{d^{\mu}_{m}}{\o^{m + \frac{1}{2}}}
\ \ \ \ ;\ \ \ \ 
\psi^{\mu}_{(+)}(\bar{\o}) = \frac{i^{-\frac{1}{2}}}{\sqrt{2\pi}}
\,\sum_{m \in {\bf{Z}}}
\frac{\tilde{d}^{\mu}_{m}}{\bar{\o}^{m + \frac{1}{2}}}.
\end{eqnarray}
Now the BC(s) (\r{28}) in case of free open super strings 
($\mathcal{B}_{\mu\nu} = 0$) requires
$d = \tilde{d}$ in the expansions (\r{14}). The expressions (\r{14})
can be equivalently written as:
\begin{eqnarray}
\label{15}
d^{\mu}_{m} &=& \frac{\sqrt{2\pi}}{\sqrt{i}}\oint \frac{d\o}{2\pi i}\, 
\o^{m - \frac{1}{2}}\, \psi^{\mu}_{(-)}(\o) 
 = \,-\sqrt{2\pi}\sqrt{i}\oint \frac{d\bar{\o}}{2\pi i}\, 
\bar{\o}^{m - \frac{1}{2}}\, \psi^{\mu}_{(+)}(\bar{\o}).
\end{eqnarray}
The anticommutation relation between $d$'s can be worked out
from the contour arguement \cite{pol} and the operator product expansion (OPE) 
(\r{11}) (with $\mathcal{B_{\mu\nu}}=0$):
\begin{eqnarray}
\label{16}
\left\{d^{\mu}_{m}, d^{\nu}_{n}\right\} &=& \frac{1}{i}
\oint \frac{d\o_2}{2\pi i} \mathrm{Res}_{\o_1 \to \o_2}\left(
\o^{m - \frac{1}{2}}_1\,  \psi^{\mu}_{(-)}(\o_1)\,
\o^{n - \frac{1}{2}}_2\,  \psi^{\nu}_{(-)}(\o_2)\right) \no \\
&=& 2\pi\alpha^{\prime}\,\eta^{\mu \nu}\, \delta_{m+n,0}
=\eta^{\mu \nu}\, \delta_{m+n,0} 
\end{eqnarray}
where we have set $2\pi\alpha^{\prime}=1$.
The anti-commutation relations between $\psi^{\mu}_{(-)}(\o, \bar{\o})$ and 
$\psi^{\nu}_{(+)}(\o^{\prime}, \bar{\o^{\prime}})$ are then obtained by using
(\r{16}):
\begin{eqnarray}
\left\{\psi^{\mu}_{(-)}(\o, \bar{\o}), \psi^{\nu}_{(-)}(\o^{\prime}, 
\bar{\o^{\prime}})\right\} &=& \frac{i\eta^{\mu \nu}}{2 \pi}
 \sum_{m \in {\bf{Z}}}\left(\o^{-m - \frac{1}{2}}\,
 \o^{\prime m - \frac{1}{2}} \right) \no \\
\left\{\psi^{\mu}_{(+)}(\o, \bar{\o}), \psi^{\nu}_{(+)}(\o^{\prime}, 
\bar{\o^{\prime}})\right\} &=& -\frac{i\eta^{\mu \nu}}{2 \pi}
 \sum_{m \in {\bf{Z}} }\left(\bar{\o}^{-m - \frac{1}{2}}\,
 \bar{\o}^{\prime m - \frac{1}{2}} \right)\no \\
\left\{\psi^{\mu}_{(-)}(\o, \bar{\o}), \psi^{\nu}_{(+)}(\o^{\prime}, 
\bar{\o^{\prime}})\right\} &=& \frac{\eta^{\mu \nu}}{2 \pi}
 \sum_{m \in {\bf{Z}}}\left(\o^{-m - \frac{1}{2}}\,
 \bar{\o}^{\prime m - \frac{1}{2}} \right).
\label{17}
\end{eqnarray}
To obtain the usual equal time ($\tau = \tau^{\prime}$)
anticommutation relation we first rewrite (\r{17}) in ``$z$ frame''
using (\r{10}, \r{13}) and then in terms of $\s^1, \ \s^2$ to find:
\begin{eqnarray}
\left\{\psi^{\mu}_{(-)}(\s^1, \s^2), \psi^{\nu}_{(-)}(\s^{1 \prime}, 
\s^{\prime 2})\right\} &=& \frac{\eta^{\mu \nu}}{2 \pi}
 \sum_{m \in {\bf{Z}}}\left[\exp\left({im(\s^1 + i \s^2
- \s^{\prime 1} - i \s^{\prime 2})} \right)\right]\no \\
\left\{\psi^{\mu}_{(+)}(\s^1, \s^2), \psi^{\nu}_{(+)}(\s^{1 \prime}, 
\s^{\prime 2})\right\} &=& \frac{\eta^{\mu \nu}}{2 \pi}
 \sum_{m \in {\bf{Z}}}\left[\exp\left({im(\s^1 - i \s^2
- \s^{\prime 1} + i \s^{\prime 2})} \right)\right]\no \\
\left\{\psi^{\mu}_{(-)}(\s^1, \s^2), \psi^{\nu}_{(+)}(\s^{1 \prime}, 
\s^{\prime 2})\right\} &=& \frac{\eta^{\mu \nu}}{2 \pi}
 \sum_{m \in {\bf{Z}}}\left[\exp\left({im(\s^1 + i \s^2
+ \s^{\prime 1} - i \s^{\prime 2})} \right)\right].
\label{18}
\end{eqnarray}
Finally substituting $\tau = \tau^{\prime}$ (i.e. $\s^2 = \s^{2\,\prime}$)
and $\s^1 = \s$ we get
back the equal time anti-commutation relations:
\begin{eqnarray}
\left\{\psi^{\mu}_{(-)}(\s, \tau), \psi^{\nu}_{(-)}(\s^{\prime}, 
\tau)\right\} &=&\eta^{\mu \nu}
 \delta_{P}(\s - \s^\prime)\no \\
\left\{\psi^{\mu}_{(+)}(\s, \tau), \psi^{\nu}_{(+)}(\s^{\prime}, 
\tau)\right\} &=& \eta^{\mu \nu}
 \delta_{P}(\s - \s^\prime)\no \\
\left\{\psi^{\mu}_{(-)}(\s, \tau), \psi^{\nu}_{(+)}(\s^{\prime}, 
\tau)\right\} &=& \eta^{\mu \nu}
 \delta_{P}(\s + \s^\prime).
\label{19}
\end{eqnarray}
where, $\delta_{P}(\s - \s^\prime)$ is the so called periodic delta function
which is defined as:
\begin{eqnarray}
\delta_{P}(\s - \s^\prime) = \frac{1}{2 \pi}\sum_{m \in {\bf{Z}}}
\exp\left(im(\s - \s^{\prime})\right). 
\label{20}
\end{eqnarray}
This structure of anticommutator is completely consistent with the
BCs (\r{28}) for ${\mathcal{B}}_{\mu\nu} = 0$. 
Note that not
only the usual Dirac delta function is replaced by the periodic delta
function but also the anticommutator among $\psi_{(-)}, \psi_{(+)}$ are
non-vanishing even in case of the free open fermionic string
\cite{agh, jing}.
\noindent The most important feature of the above analysis is that unlike
the bosonic case, the new normal ordering of the fermionic 
operators (that incorporates the BC(s)) (\ref{11}) leads to the
nonanticommutative structures (\ref{19}) among the fermionic 
string coordinates.

\subsection{Open superstring in the constant ${\mathcal{B}}$-field background} 
We now analyse the open superstring moving in presence of a background
antisymmetric tensor field ${\cal{B}}_{\mu \nu}$. 
To begin with, let us again consider the Laurent expansion of 
$\psi_{(-)}^{\mu}(\o)$ and $\psi_{(+)}^{\mu}(\bar{\o})$ 
(\r{14}). Now due to the BC(s) (\r{28}) (with $\mathcal{B}_{\mu \nu} \neq 0$),
the modes $d$ and $\tilde{d}$ are no longer
independent but satisfy the following relation:
\begin{eqnarray}
E_{\mu \nu}\, d^{\nu}_{m}\, = \, E_{\nu \mu}\, \tilde{d}^{\nu}_{m}. 
\label{21}
\end{eqnarray}
Hence there exists only one set of independent modes 
$\alpha^{\mu}_{m}$, which can be thought of as the modes of free open
strings and is related to $d^{\mu}_{m}$ and $\tilde{d}^{\mu}_{m}$ by: 
\begin{eqnarray}
\label{22}
d^{\mu}_{m} &=& \left(\delta^{\mu}_{\ \nu} - {\cal{B}}^{\mu}_{\ \nu}
\right)\alpha^{\nu}_{m} := \left[({1\!\mbox{l}} - {\cal{B}})\alpha
\right]^{\mu}_{m} \no \\
\tilde{d}^{\mu}_{m} &=& \left(\delta^{\mu}_{\ \nu} + {\cal{B}}^{\mu}_{\ \nu}
\right)\alpha^{\nu}_{m} := \left[({1\!\mbox{l}} +
{\cal{B}})\alpha\right]^{\mu}_{m} .
\end{eqnarray}
Note that under world-sheet parity transformation 
(i.e. $\sigma\leftrightarrow-\sigma$), 
$d^{\mu}_{m} \leftrightarrow \tilde{d}^{\mu}_{m}$, since ${\cal{B}}_{\mu \nu}$
is a world-sheet pseudo-scalar (similar to bosonic part \cite{our}).  
Substituting (\r{22}) in (\r{14}), we obtain  the 
following Laurent expansions for $\psi^{\mu}_{-}$ and $\psi^{\mu}_{+}$:
\begin{eqnarray}
\label{23}
\psi^{\mu}_{(-)}(\o) &=& \frac{i^{\frac{1}{2}}}{\sqrt{2\pi}}
\sum_{m \in {\bf{Z}}} 
\frac{\left[({1\!\mbox{l}} -
{\cal{B}}) \alpha\right]^{\mu}_{m}}{\o^{m + \frac{1}{2}}} \\
\psi^{\mu}_{(+)}(\bar{\o}) &=& \frac{i^{- \frac{1}{2}}}{\sqrt{2\pi}}
\sum_{m \in {\bf{Z}}}
\frac{\left[({1\!\mbox{l}} + {\cal{B}})
\alpha\right]^{\mu}_{m}}{\bar{\o}^{m + \frac{1}{2}}}\no.
\end{eqnarray}
These are the appropriate mode expansions for the fermionic part of the
interacting superstring,
that satisfy both the equations of motion (\ref{6}) and the BC(s) (\ref{28}).

\noindent Now the expressions (\r{23}) for interacting superstrings can 
also be written as:
\begin{eqnarray}
\left[({1\!\mbox{l}} - {\cal{B}})\alpha\right]^{\mu}_{m} &=& 
\frac{\sqrt{2\pi}}{i}\oint \frac{d\o}{2\pi i}\, 
\o^{m - \frac{1}{2}}\, \psi^{\mu}_{(-)}(\o) \no \\
\left[({1\!\mbox{l}} + {\cal{B}})\alpha\right]^{\mu}_{m} &=& 
\,\frac{\sqrt{2\pi}}{i}\oint \frac{d\bar{\o}}{2\pi i}\, 
\bar{\o}^{m - \frac{1}{2}}\, \psi^{\mu}_{(+)}(\bar{\o}) .
\label{24}
\end{eqnarray}
The anticommutation relation between $\alpha$'s can be obtained once again
from the contour arguement (using (\r{24})) and the $\psi\, \psi$ OPE (\r{11}):
\begin{eqnarray}
\label{25}
\left\{\alpha^{\mu}_{m}, \alpha^{\nu}_{n}\right\} 
=  \left[\left({1\!\mbox{l}} - {\cal{B}}^2
\right)^{-1}\right]^{\mu \nu} \delta_{m, -n}
= \left(\mathcal{M}^{-1}\right)^{\mu \nu} \delta_{m, -n} 
\end{eqnarray}
where, $\mathcal{M} = ({1\!\mbox{l}} - {\cal{B}}^2)$ ; $({\cal{B}}^2)^{\mu \nu} 
= {\cal{B}}^{\mu}_{\ \rho}{\cal{B}}^{\rho \nu}$\footnote{Here we 
should note that $\left({1\!\mbox{l}}\right)^{\mu \nu} 
= \eta^{\mu \nu}$.}. 
Now the anticommutator between the fermionic string coordinates 
can be computed using (\r{23}), (\r{25}). The antibrackets between
$\left\{\psi^{\mu}_{(-)}, \psi^{\nu}_{(-)}\right\}$ and 
$\left\{\psi^{\mu}_{(+)}, \psi^{\nu}_{(+)}\right\}$ are the same
as that of free case but the anticommutator between 
$\psi^{\mu}_{(-)}$ and $\psi^{\mu}_{(+)}$ gets modified to the following form:
\begin{eqnarray}
\left\{\psi^{\mu}_{-}(\o, \bar{\o}), \psi^{\nu}_{+}(\o^{\prime}, 
\bar{\o^{\prime}})\right\} &=& \frac{1}{2 \pi} \, 
 \sum_{m \in {\bf{Z}}} \left[\frac{\left({1\!\mbox{l}} - {\cal{B}}
\right)^{\mu}_{\ \rho}\, \left[\left({1\!\mbox{l}} - {\cal{B}}^2\right)^{-1}
\right]^{\rho \s}\,
\left({1\!\mbox{l}} - {\cal{B}}\right)_{\s}^{\ \nu}
}{\o^{m + \frac{1}{2}}\, \bar{\o}^{\prime -m + \frac{1}{2}}}\right].
\label{26}
\end{eqnarray}
Now proceeding as before, we can write the above anticommutation relation
in $(\tau, \s)$ coordinates to obtain the usual equal time
(i.e. $\tau = \tau^{\prime}$) anticommutation relation:
\begin{eqnarray}
\left\{\psi^{\mu}_{(-)}(\s, \tau), \psi^{\nu}_{(+)}(\s^{\prime}, 
\tau)\right\} &=& E^{\rho\, \mu}\, 
\left[\left({1\!\mbox{l}} - {\cal{B}}^2\right)^{-1}
\right]_{\rho\, \s}\, E^{\nu \,\s}\,
 \delta_{P}(\s + \s^\prime).
\label{27}
\end{eqnarray}
The above result reduces to the the free case result in the 
$\mathcal{B}_{\mu\nu}=0$ limit and also agrees with the existing results
in the literature \cite{agh}, \cite{jing}.


\section{Conclusions}
In this paper, we have used conformal field theoretic techniques
to compute the anticommutator among Fourier components of fermionic sector
of super strings. Using this the anticommutator between the
basic fermionic fields is obtained. 
This is the extension of our earlier work on bosonic strings
\cite{our}. The method is also different from (\cite{jing}), 
where the algebra among the Fourier components
have been computed using the Faddeev-Jackiw symplectic formalism.
The advantage of this approach (as mentioned in 
\cite{our} also) is that the results one obtains takes
into account the quantum effects right from the beginning,
in contrary to the previous 
investigations, which were made essentially at the classical level 
\cite{chu, chu1, rb, jing, br1}. Interestingly, 
the new normal ordering that takes into account the
effect of the BC(s) plays a crucial role in obtaining the
nonanticommutative symplectic structure among the fermionic string
coordinates (\r{19}). This is in contrast to the analysis in case of the bosonic
strings where the new normal ordering has no bearing on the symplectic
structure. 
\noindent Finally, we also computed the oscillator 
algebra in presence of the $\mathcal{B}$
field which is a parity-odd field on the string world-sheet. As in the
bosonic case,
in presence of this $\mathcal{B}$ field, 
the fourier modes appearing in the  Laurent series expansions 
of the fermionic fields $\psi^{\mu}_{(-)}$ and $\psi^\mu_{(+)}$  
of the closed string are 
no longer equal when open string BCs (\r{28}) are imposed. These rather get
related to the free oscillator
modes $d^{\mu}_{m}$.
Using these expressions of the modes (\ref{22}), we rewrite
the fermionic fields $\psi^{\mu}_{(-)} $ and $\psi^{\mu}_{(+)}$
entirely in terms
of the free oscillator modes $\alpha^{\mu}_{m}$ (\r{22}). Then
a straight forward calculation, involving $\psi\psi$ OPE (\ref{11})
and contour arguement yields the NC anticommutator given in (\r{27}), 
thereby reproducing the results of \cite{agh}. 
\section*{Appendix}
Here we would like to give some of the
computational details involved in deriving 
(\ref{eqmov2}) from (\ref{8}) and (\ref{28}) (for convenience we treat the
free case, i.e. $\mathcal{B} = 0$). 
Eq.(\ref{8}) with $z$ replaced by $\omega$ yields:
\begin{eqnarray}
\label{a1}
0&=&\int [d\psi]\left[  \frac{\delta}{\delta \psi^{\mu} _{(a)}
(\omega,\bar{\omega})}[e^{-S_{F}} \psi^{\nu}_{(b)} (\omega', \bar{\omega}')]
  \right] \nonumber\\
&=&\int[d\psi]e^{-S_{F}}
\left[-\frac{\delta S_{F}}{\delta\psi_{(a)}(\omega,\bar{\omega})}
\psi_{(b)}(\omega^{\prime},\bar{\omega}^{\prime})+
\frac{\delta\psi_{(b)}(\omega^{\prime},\bar{\omega}^{\prime})}
{\delta\psi_{(a)}(\omega,\bar{\omega})}\right]
\end{eqnarray}
Putting $a=+$, $b=-$; we obtain:
\begin{eqnarray}
\label{a2}
0&=&\int[d\psi]e^{-S_{F}}
\left[\frac{i}{2\pi\alpha^{\prime}}\partial_{\omega}\psi_{+}
(\omega,\bar{\omega})\psi_{-}(\omega^{\prime}, \bar{\omega}^{\prime})+
\frac{\delta\psi_{(-)}(\omega^{\prime},\bar{\omega}^{\prime})}
{\delta\psi_{(+)}(\omega,\bar{\omega})}\right] 
\end{eqnarray}
where we have dropped the boundary term (arising from
the first term in (\ref{a1})) as it vanishes due to the
BC(s) (\ref{28}) (with $\mathcal{B}=0$).\\

\noindent Now we discuss two distinct cases seperately.

\noindent $\bullet$ Case 1: The insertion 
$\psi_{(-)}(\omega^{\prime},\bar{\omega}^{\prime})$ is not located 
at the boundary:

\noindent In this case 
\begin{eqnarray}
\label{a3}
\frac{\delta\psi_{(-)}(\omega^{\prime},\bar{\omega}^{\prime})}
{\delta\psi_{(+)}(\omega,\bar{\omega})}=0
\end{eqnarray}
and therefore one finds:
\begin{eqnarray}
\label{a4}
\langle \partial_{\o} \psi^{\mu}_{(+)} (\o,\bar{\o}) \psi^{\nu}_{(-)}
(\o',\bar{\o}') \rangle &=& 0.
\end{eqnarray}

\noindent $\bullet$ Case 2: The insertion 
$\psi_{(-)}(\omega^{\prime})$ is located at the boundary (since
$\o' = \bar{\o}'$ at the boundary, the insertion 
$\psi_{(-)}(\omega^{\prime})$ depends only on the arguement $\o'$): \\
In this case the computation of the second term in (\ref{a2})
needs to be done more carefully. One finds
\begin{eqnarray}
\label{a5}
\frac{\delta\psi_{(-)}(\omega^{\prime},\bar{\omega}^{\prime})}
{\delta\psi_{(+)}(\omega,\bar{\omega})}\Big{\vert}_{\o' = \bar{\o}'} &=& 
i\, \frac{\delta\psi_{(+)}(\omega^{\prime},\bar{\omega}^{\prime})}
{\delta\psi_{(+)}(\omega,\bar{\omega})}\Big{\vert}_{\o' = \bar{\o}'}
\no \\
&=& i\, \delta^2 \left(\o-\o', \bar{\o}- \bar{\o}'\right)\Big{\vert}_{\o' = \bar{\o}'} \no \\ 
&=&  i\, \delta^2 \left(\o-\o', \bar{\o}- \o'\right).
\end{eqnarray}
where we have used the BC (\ref{28}) (with $\o$ repaced by $\o^{\prime}$)
in the first line of (\ref{a5}).

\noindent Substituting (\ref{a5}) in (\ref{a2}) and equating the volume term to zero,
one finds the third of the equations in (\ref{eqmov2}) 
(with $\mathcal{B} = 0$). 

\noindent
Similarly, for other choices of $a, b$
the rest of the equations in (\ref{eqmov2}) can be derived (with 
$\mathcal{B} = 0$).

\section*{Acknowledgment}
The authors would like to thank the referee for very useful comments.


\end{document}